# Thermal generation of spin current in epitaxial $CoFe_2O_4$ thin films

Er-Jia Guo,[1,a] Andreas Herklotz,[2] Andreas Kehlberger,[1] Joel Cramer,[1] Gerhard Jakob,[1] and Mathias Kläui [1,a]

[1] *Institut für Physik, Johannes Gutenberg-Universität Mainz, 55099 Mainz, Germany*

[2] *Materials Science and Technology Division, Oak Ridge National Laboratory, Oak Ridge, Tennessee 37831, USA*

**Abstract**

The longitudinal spin Seebeck effect (LSSE) has been investigated in high-quality epitaxial $CoFe_2O_4$ (CFO) thin films. The thermally excited spin currents in the CFO films are electrically detected in adjacent Pt layers due to the inverse spin Hall effect (ISHE). The LSSE signal exhibits a linear increase with increasing temperature gradient, yielding a LSSE coefficient of ~100 nV/K at room temperature. The temperature dependence of the LSSE is investigated from room temperature down to 30 K, showing a significant reduction at low temperatures, revealing that the total amount of thermally generated magnons decreases. Furthermore, we demonstrate that the spin Seebeck effect is an effective tool to study the magnetic anisotropy induced by epitaxial strain, especially in ultrathin films with low magnetic moments.

---

a) Author to whom correspondence should be addressed.
Electronic mail: ejguophysics@gmail.com, and klaeui@uni-mainz.de

The spin Seebeck effect (SSE) refers to the conversion of thermal excitations of magnetic orders into a spin current, which can be pumped into an adjacent non-magnetic metal layer and further be converted into electric voltage due to the inverse spin Hall effect (ISHE). This effect was firstly discovered in permalloy films and subsequently in the ferromagnetic semiconductor GaMnAs and in the Heusler compound $Co_2MnSi$.[1-3] Furthermore, the same effect was observed in the ferrimagnetic insulators $LaY_2Fe_5O_{12}$ and $Y_3Fe_5O_{12}$ (YIG) single crystals,[4-5] which opened up an extremely exciting research topic called "*insulator spintronics*". In magnetic insulators, there are no mobile charges involved, instead, the excitation of the localized spin-polarized electrons is the primary source of the spin currents. This feature stimulates intensive research on such magnetic insulators due to its potential advantages in greatly reducing the power dissipation compared with conventional semiconductor-based thermal electric devices.[6-7] Up to now, the SSE has been investigated only for a limited number of insulating ferrimagnets. Most researches focused on garnet ferrites, like YIG, due to their ultralow damping constants (in the order of $10^{-4}$) and the relative high Curie temperatures of $T_C$ ~550 K.[4, 5, 8-13] Current studies on YIG have revealed that the interface effects plays an important role for the detected ISHE signals.[13] The complexity of the garnet ferrites' lattice structure makes the interface termination with adjacent detection layers complicated. Therefore, there is a key need to explore the insulating SSE materials which possess relatively simple lattice structures with a view to use them for reliable SSE systems.

Spinel ferrites are one of the promising SSE candidates, since they provide a less complex crystal structure compared to YIG. In previous works, the observation of the SSE in a ferrimagnetic polycrystalline $NiFe_2O_4$ μm-thick film[14] and a sintered polycrystalline $(Mn,Zn)Fe_2O_4$ slab[15] has been reported. However, the intrinsic semiconducting property of the polycrystalline samples led to the coexistence of the anomalous Nernst effect (ANE) and the SSE in the detected ISHE signals. By replacing the cations on the octahedral B-sites with $Co^{2+}$, another representative spinel ferrite, cobalt-ferrite ($CoFe_2O_4$, CFO), can be prepared. It is a ferrimagnetic oxide with a Curie temperature well above room temperature ($T_C$ ~790 K) and is an excellent insulator from room temperature down to low temperatures,[16-18] providing an ideal system to study the SSE over a wide temperature range. Very recently, Isasa *et al*.[19] have reported the observation of spin Hall magnetoresistance (SMR) in Pt layers in conjunction with CFO thin films. However, other relevant spin-dependent phenomena have not been explored so far. In this paper, we report the observation of the longitudinal SSE in high-quality epitaxial CFO films and reveal its dependences on the temperature and the magnetic field. Furthermore, we demonstrate that the SSE is an effective tool to probe magnetic properties of a single low magnetic moment film. Using the SSE is particularly attractive for systems with films of small magnetic moment and magnetic substrates, where conventional measurements are challenging due to dominating substrate contributions.

CFO thin films with thicknesses of 20 nm and 168 nm were epitaxially grown on (001)-oriented MgO substrates by pulsed laser deposition (PLD), respectively. Sintered ceramic CFO targets were ablated with a KrF excimer laser pulse with a wavelength of 248 nm and a repetition frequency of 5 Hz. During the deposition, the temperature of the substrates was kept at 660° C with an oxygen pressure of 0.07 mbar. Afterwards, the films were *in-situ* annealed in the pure oxygen with a pressure of 0.5 *atm* for 15 mins and then slowly cooled down to room temperature. By measuring the x-ray reflectivity (XRR), the thickness of the 20 nm-thick CFO films was determined, while the thickness of the thicker films was estimated from the deposition time. The crystalline quality of the films was examined by using a high-resolution x-ray diffractometer (HRXRD, Bruker D8-DISCOVER). The rocking curves of the CFO films show a high crystalline quality with a full width at half maximum (FWHM) of less than 0.01°. Figure 1(a) shows the representative $\omega$-$2\theta$ scans for CFO thin films. The clear Laue oscillations reveal the high uniformity of the films. The atomic force microscopy (AFM) image in Fig. 1(b) reveals a smooth surface of the as-grown 168 nm-thick CFO films with an rms roughness of 0.28 nm. The strain states of two CFO films are examined by performing reciprocal space mapping (RSM) [Fig. 1(c) and (d)]. The area scans show that all the CFO films are coherently grown on the MgO substrates, indicating that the CFO films have the same in-plane lattice parameter ($a$) as the substrate. The bulk lattice parameters of CFO and MgO are $a_{CFO}$ = 8.39 Å and $a_{MgO}$ = 4.21 Å, respectively. Thus, the CFO films are grown under an in-plane tensile strain. The out-of-plane lattice parameters ($c$) are determined to be 8.34 Å and 8.37 Å for 20 nm- and 168 nm-thick CFO films, respectively. We find that the tetragonal distortion $t$ [$= \frac{2(a-c)}{a+c}$] of the CFO films reduces from 1% to 0.68% as the film thickness increases from 20 nm to 168 nm.

To conduct the SSE measurements, a thin Pt pattern with a thickness of 5 nm is sputtered on top of the CFO films through a shadow mask. Pt is chosen as spin detector due to its high spin Hall angle.[8, 9, 11-15] As illustrated in Fig. 2(a), the Pt stripes A and B are oriented along the [0$\bar{1}$0] and [$\bar{1}$00] in-plane direction, respectively. Both stripes have the same geometry with a length of $L_{pt}$ = 6 mm and a width of $w_{pt}$ = 1 mm. Our SSE measurements are carried out using the longitudinal configuration with the temperature gradient parallel to the surface normal. The whole setup is mounted onto a copper heat sink and put into a cryostat with a variable temperature insert. A resistive heater is glued with high thermally conducting grease on top of the Pt stripes. The thermal gradient is created by applying a heat current ($I_{heat}$) through the resistive heater. Thus, the spin currents can be thermally generated in the CFO films and injected into the Pt layer across the interface. If the magnetization ($\vec{M}$) of the CFO layer is aligned along the $y$ (or $x$) direction, an electric field ($\vec{E}_{ISHE}$) is generated along the Pt stripe A (or B) according to the relation,[1, 2]

$$\vec{E}_{ISHE} \propto \vec{J}_s \times \vec{\sigma} \qquad (1)$$

where $\vec{J}_s$ and $\vec{\sigma}$ denote the spatial direction of the spin current and the spin polarization vector parallel to $\vec{M}$. $\vec{E}_{ISHE}$ can be electrically detected by measuring the voltage drop ($V_{SSE}$) at the two ends of the Pt stripes.

We start with applying a fixed temperature gradient to the Pt/CFO sample by fixing $I_{heat}$ to 13 mA and recording $V_{SSE}$ as a function of the swept magnetic field. Figure 2(b) shows the field dependent $V_{SSE}$ for stripe A and B in the 168 nm-thick CFO/Pt sample at 300 K. The in-plane magnetic field is applied perpendicular to the long axis of the stripes in order to provide the maximum value of $\vec{E}_{ISHE}$ according to Eq. (1). $V_{SSE}$ flips its sign when the direction of the magnetic field reverses, reflecting the magnetization reversal of the CFO film. We note that the magnitude of $V_{SSE}$ for stripe A and B are the same within the error range. The in-plane magnetocrystalline anisotropy of the cubic (or tetragonally-distorted) spinel structure is expected to exhibit a fourfold symmetry so that the ISHE signals in the two stripes should be identical. We also measure the $V_{SSE}$-$H$ curves of stripe A and B for out-of-plane magnetic fields, as shown in Fig. 2(d). No signal is detected since the spins of the electrons are oriented parallel to the spatial direction of the spin current, resulting in a zero $E_{ISHE}$. The field dependent magnetization of the CFO films is measured by a superconducting quantum interference device (SQUID, Quantum Design) magnetometer. Figure 2(c) and (e) show the $M$-$H$ hysteresis loops at 300 K for the in-plane magnetic fields along the [010] and [$\bar{1}$00] direction and the out-of-plane magnetic fields, respectively. The saturated magnetization ($M_S$) in both, in-plane and out-of-plane field direction, gives nearly the same value of ~195 emu/cm$^3$ which is a usual value observed in CFO films.[16-20] The coercive field ($H_C$) for an in-plane field (<10 mT) is significantly smaller than that for an out-of-plane field (180 mT). Also, the out-of-plane remanent magnetization ($M_R$) is considerable larger than in-plane. This illustrates the remarkably large out-of-plane magnetic anisotropy that is a result of the tensile strain and the extraordinary high negative magnetostriction of CFO due to the presence of unquenched orbital momentum in $Co^{2+}$ cations.[19, 20] We find the shape of $V_{SSE}$ reflects the behavior of $\vec{M}$, suggesting that the observed voltages clearly originate from the magnetic nature of the CFO films.

The thermal gradient can be varied by changing the $I_{heat}$ applied to the resistive heater. Figure 3(a) presents the field dependent room-temperature $V_{SSE}$ for the 168 nm-thick CFO/Pt sample with $I_{heat}$ increasing from 0 to 13 mA. With increasing $I_{heat}$ the saturated value of $V_{SSE}$ increases correspondingly, whereas there is no visible signal observed at $I_{heat}$ = 0 mA. The exact temperature differences ($\Delta T$) across the Pt/CFO bilayer can be calculated by utilizing the Pt stripes on the top and bottom surfaces of the samples as resistance-temperature sensors. In Fig. 3(b), we plot the saturated $V_{SSE}$ values as a function of $\Delta T$ at 300 K. We find that the $V_{SSE}$ signals are linearly proportional to $\Delta T$, yielding the SSE coefficient $\sigma_{SSE}$ (= $V_{SSE}/\Delta T$) of (94.1 ± 3.2) nV/K at 300 K. Similar room-temperature SSE coefficient values have been reported in other spinel oxides. Meier *et al.*[14] found a SSE coefficient of ~30 nV/K for $\mu$m-thick polycrystalline $NiFe_2O_4$ films at 300 K. Uchida *et al.*[15] measured a SSE coefficient of ~200 nV/K for a polycrystalline $(Mn,Zn)Fe_2O_4$ slab at room temperature. However, we notice that the $\sigma_{SSE}$ of CFO films is nearly five times smaller than that of PLD grown YIG films of similar thickness although the saturation magnetizations of two materials are approximately the same.[21] The large deviation may be attributed to the differences in the Gilbert damping of the bulk materials or the interface qualities of different samples.

The field-sweep SSE measurements are also conducted at various temperatures. Figure 3 (b) shows the $V_{SSE}$-$\Delta T$ curves measured at 50 K and 100 K. Analogous to the room temperature measurement, $V_{SSE}$ increases linearly with increasing $\Delta T$, verifying that the detected signals originate from the ISHE signals in the Pt layer. By fitting the experimental results, we observe that the $\sigma_{SSE}$ reduces from (74.5 ± 2.6) nV/K at 100 K to (53.8 ± 1.1) nV/K at 50 K. Continuous temperature dependent $\sigma_{SSE}$ measurements are carried out using a temperature-sweep method, at which the magnetic field is kept constant at ±7 T while the temperature is swept twice (first at +7 T and then at -7 T) from 300 K down to 30 K. The temperature is swept with a very slow speed of -0.25 K/min in order to keep the thermal equilibrium between the sample and the cryostat during the temperature drop. The SSE coefficients are determined by $\sigma_{SSE} = [V_{SSE}(7T)-V_{SSE}(-7T)]/2\Delta T$. Figure 3(c) shows $\sigma_{SSE}$ as a function of temperature for the 20 nm- and 168 nm-thick CFO/Pt sample. Results obtained from field-sweep measurements are marked as open symbols in Fig. 3(c). The good agreement between two measuring methods validates the reproducibility and the reliability of our results. Comparing the signal amplitudes, $\sigma_{SSE}$ of the 20 nm-thick CFO films is much smaller than that of the 168 nm-thick CFO films in particular at high temperatures. For both films $\sigma_{SSE}$ stays nearly constant at high temperatures and shows a significant decrease at low temperatures. Similar temperature dependences have been reported in both, NiFe$_2$O$_4$ films[14] and YIG thin films with thickness below 600 nm.[21-25] Our results can be interpreted based on the concept of the magnonic flux in the CFO films. The amplitude of $\sigma_{SSE}$ is determined by the number of thermally excited magnons which can reach the Pt/CFO interface and further be converted into the ISHE signal. In our previous work, we revealed a finite magnonic propagation length in YIG bulk materials.[21] This universal mechanism can also be used to interpret our results observed in the CFO films. For films with thicknesses smaller than the magnonic propagation length, the total number of thermally excited magnons in a thinner film is less than that of a thicker film exposed to the same temperature gradient. Hence a smaller $\sigma_{SSE}$ is expected for the thinner CFO films. The magnons are the bosonic quasi-particles and their number is described by the magnonic density of states and followed the Bose-Einstein distribution. The best approximation for the number of thermally excited magnons is given by the experimentally accessible heat capacitance of the magnetic lattice. For low temperatures the thermal conductivity shows a remarkably reduction,[26, 27] yielding a rapid decrease of the total number of thermally excited magnons. This effect will lead to the reduction of the SSE signals at low temperatures. Furthermore, we also notice that there is no significant magnetic field induced suppression of the SSE signals at ±7 T, as shown in Fig. 2(b) and Fig. 3(a). This result is consistent with the field suppressed SSE measurements on the YIG films with thicknesses less than 300 nm.[22,24,25] In the previous work, we demonstrate that a significant magnetic field induced SSE suppression is only observed in the thick YIG films.[22,23] When the film thickness is larger than (or comparable with) the magnonic propagation length of the bulk materials, the amount of the thermal magnons that can reach the interface reduces due to the magnetic field induced reduction of the magnonic propagation length. Thus the detected ISHE signal is suppressed under applied magnetic field. In our work, the invisible field suppression of the SSE signal may attribute to the small thickness of the CFO films compared to the magnonic propagation length, within which the high magnetic fields can barely change the magnonic distribution in the CFO films.[22, 23] In addition, we want to point out that the resistance of CFO films have been recorded as a function of temperature, confirming the highly insulating nature of our films over the whole temperature range, in contrast to the previous results on the $NiFe_2O_4$ films.[14] Hence other thermoelectric phenomena, such as the conventional Seebeck effect and the ANE in the CFO films, can be excluded in our experiment, allowing one for an unambiguous identification of the SSE.

Finally, we demonstrate that the SSE is capable of electrically probing the in-plane magnetization of ultrathin CFO films. Figure 4 presents the comparison of the field dependent $V_{SSE}$ for 20 nm- and 168 nm-thick CFO films at 300 K. The shape of $V_{SSE}$-$H$ curves changes with reducing film thickness, which is further confirmed by measuring the $M$-$H$ curves of the same samples. This effect can be attributed to the change of the out-of-plane strain states of the CFO films. CFO is known for its large magnetic anisotropy that can be easily tuned by epitaxial strain.[16-18] As the film thickness reduces the out-of-plane strain of the CFO films relaxes, causing the magnetization to rotate towards the film plane. Precisely determination of the magnetization of ultrathin films is challenging due to their weak magnetic moment and the contribution from paramagnetic substrates. The SSE is found to be an alternative way to probe the magnetization via thermally excited magnons in the magnetic thin layers that is not affected by paramagnetic substrate contributions.

In summary, we have experimentally observed the longitudinal SSE in CFO thin films. The application of a temperature gradient perpendicular to the Pt/CFO interface leads to an ISHE voltage in the Pt layers, whose magnitude is proportional to the in-plane magnetization of the CFO films. Temperature dependent SSE measurements show that the signal decreases significantly at very low temperatures, revealing that the total amount of thermally generated magnons reduces correspondingly. Our results highlight that research on the SSE should be extended to various magnetic insulators, such as "soft" magnetic spinel ferrites or even "hard" magnetic hexagonal ferrites. These materials have some advantages over the current extensively studied garnet ferrites. First of all, the lattice constants of these structures are very close to each other. This helps to grow them epitaxially on various substrates, allowing us to design well-defined interfaces as well as giving much freedom to tune the magnetic properties via cation doping, strain engineering, and artificial designing of heterostructures. Secondly, the lattice structures of spinel ferrites are much simpler compared to those of garnets. This will simplify theoretical calculations and further help to reveal the genuine origin of spin currents and magnonic transport properties in bulk materials. Moreover, advanced fabrication methods [28] are capable of producing large-scale and low-cost spinel ferrite thin films, which is a crucial step for the potential application of these SSE materials in future.

**Acknowledgements**

This work is supported by Deutsche Forschungsgemeinschaft (DFG) SPP 1538 "Spin Caloric Transport", the Graduate School of Excellence Materials Science in Mainz (MAINZ) and the EU projects (IFOX, NMP3-LA-2012246102, INSPIN FP7-ICT-2013-X 612759.). Additionally, A.H. is supported by the U. S. Department of Energy, Office of Science, Basic Energy Sciences, Materials Science and Engineering Division in Oak Ridge National Laboratory, TN, USA.

**Figure captions**

**Figure 1**. (color online) (a) XRD $\omega$-$2\theta$ scans around (002) reflections for CFO thin films with thickness of 20 nm and 168 nm. (b) AFM topography image of CFO(168 nm)/MgO sample over an area of 4×4 $\mu m^2$. (c) and (d) Reciprocal space maps (RSMs) around the {103} peaks of the MgO substrates and the CFO films with thickness of 20 nm and 168 nm, respectively.

**Figure 2**. (color online) (a) Sketch of the experimental configuration used for longitudinal SSE measurements. The magnetic field dependent $V_{SSE}$ is measured for stripe A and B when the $H$ is applied (b) parallel to [$\bar{1}00$] and [010] in-plane direction and (d) parallel to the [001] out-of-plane direction, respectively. The $M$-$H$ curves of 168 nm thick CFO films with applied in-plane (c) and out-of-plane (e) magnetic fields. All the measurements are taken at 300 K.

**Figure 3**. (color online) (a) Room-temperature field dependent $V_{SSE}$ of 168 nm-thick CFO films for different thermal gradients. The temperature gradients can be altered by simply applying different heating currents ranging from 0 to 13 mA to the attached resistive heater. (b) The SSE voltages as a function of $\Delta T$ at fixed environmental temperatures of 50, 100, and 300 K, respectively. (c) The calculated $\sigma_{SSE}$ as a function of temperature for 20 nm and 168 nm-thick CFO films. The open symbols indicate the fitted $\sigma_{SSE}$ from magnetic field-sweep measurements.

**Figure 4**. (color online) The $V_{SSE}$ and $M$ as a function of magnetic field for 20 nm and 168 nm thick CFO films at 300 K. The magnetic field is applied along the [010] in-plane direction. The open symbols represent the $V_{SSE}$, while the solid lines represent the total magnetization $M$. The blue open symbols and line indicate the results of a 20 nm- thick CFO film and have been multiplied by two from the original data for comparison.

**FIG. 1.**

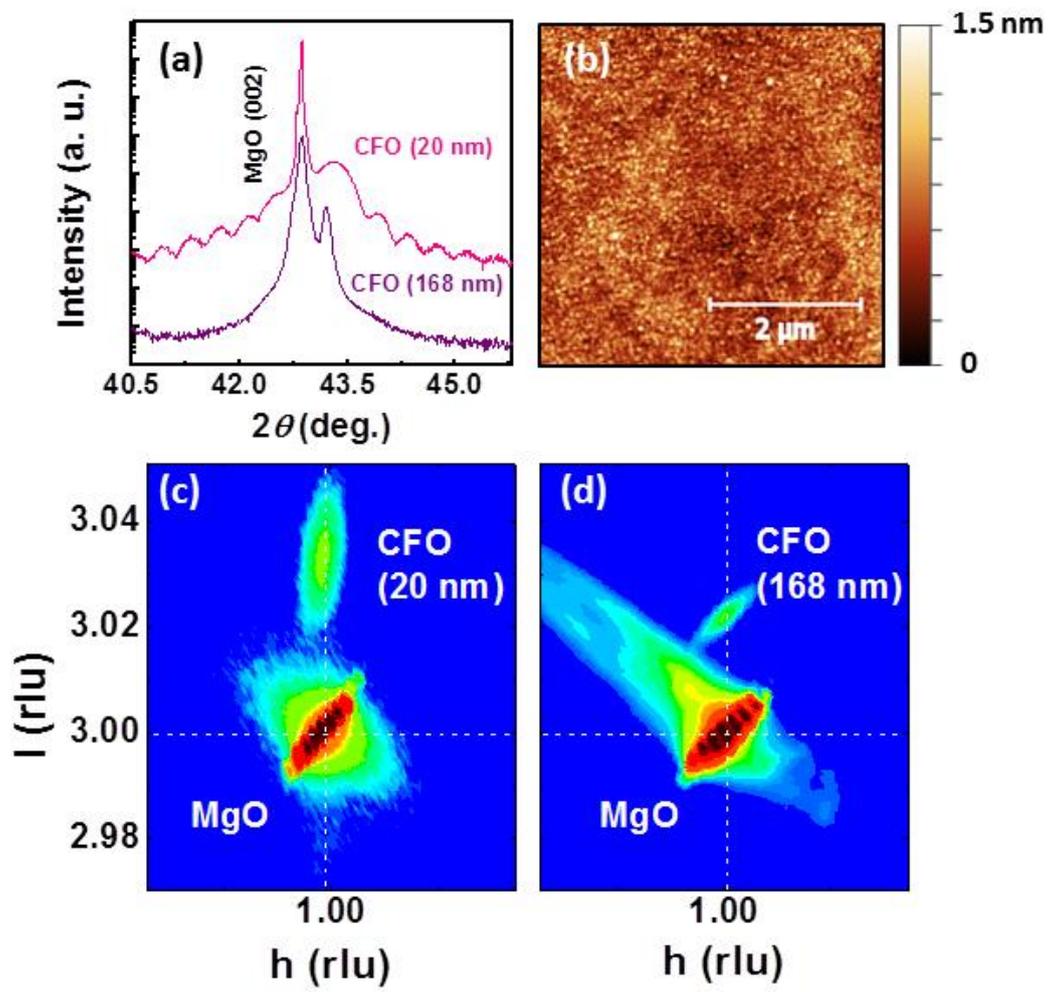

**FIG. 2.**

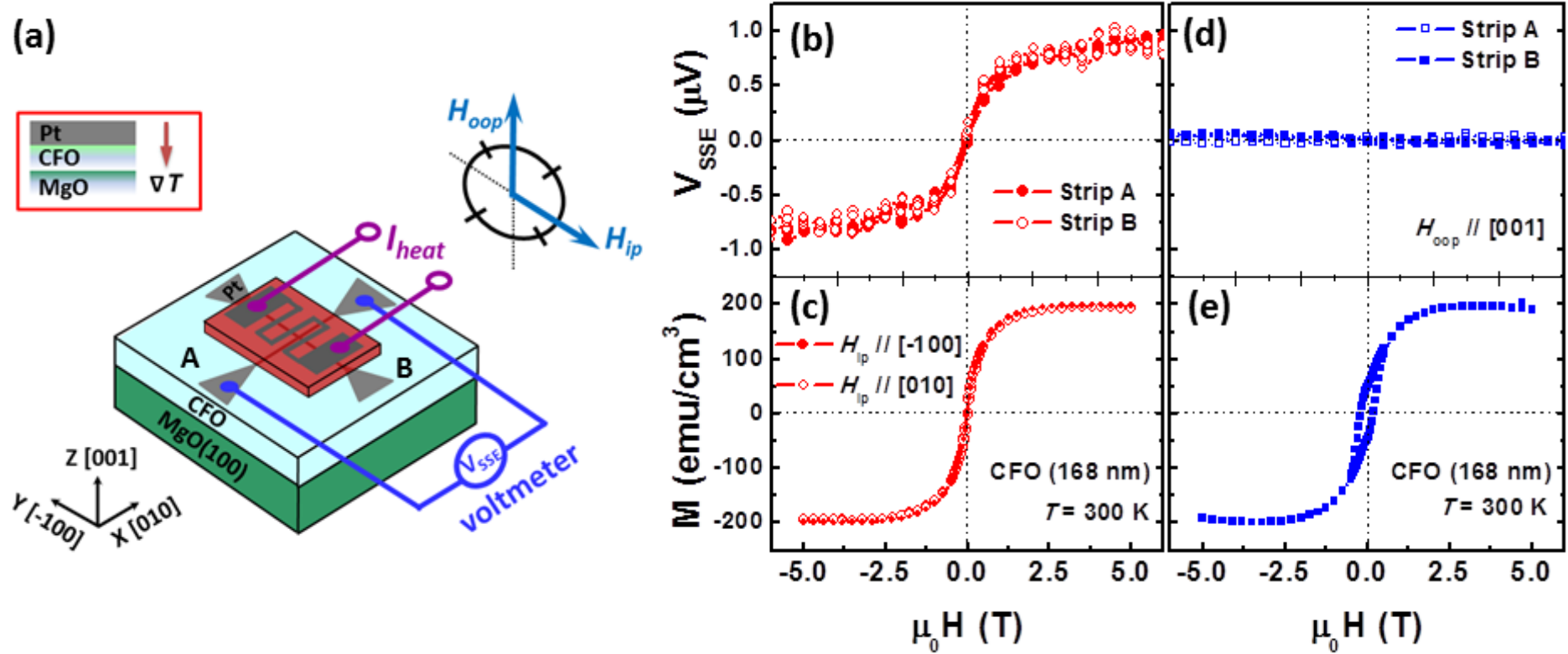

**FIG. 3.**

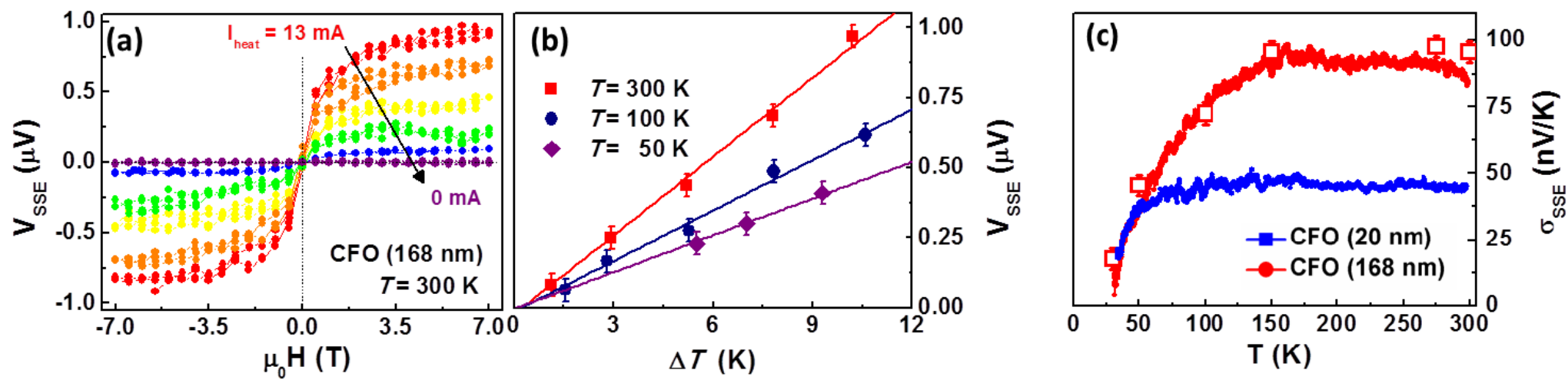

**FIG. 4.**

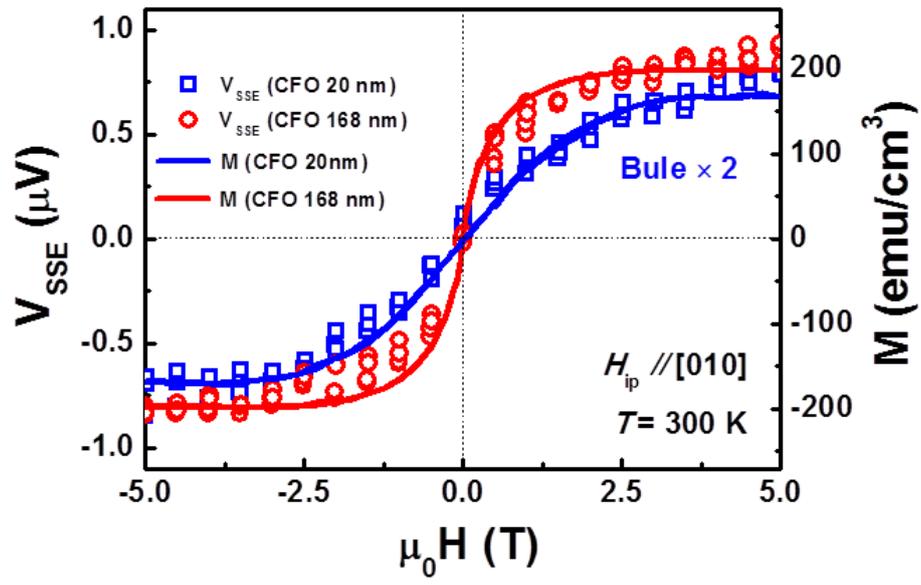